\documentclass[sigconf,screen,nonacm]{acmart}
        
\AtBeginDocument{%
  }

\usepackage{enumitem}

\newcommand{\pin}{p^{\mathrm{in}}}
\newcommand{\pout}{p^{\mathrm{out}}}
\newcommand{\pcache}{p^{\mathrm{cache}}}
\newcommand{\fix}[1]{#1}

\begin{document}

\title{TurnCost: Turns, Not Observations: Cache-Aware Loop Engineering for Coding Agents}

\author{Pramod Kumar}
\orcid{0009-0000-7426-367X}
\affiliation{%
  \institution{The Pennsylvania State University}
  \city{University Park}
  \country{USA}
}
\email{pjk5502@psu.edu}

\author{Raj Kumar Pandey}
\orcid{0009-0008-4553-3018}
\affiliation{%
  \institution{The Pennsylvania State University}
  \city{University Park}
  \country{USA}
}
\email{rzp5412@psu.edu}

\begin{abstract}
AI coding agents operate in iterative loops: they modify code, run tools, inspect the results, and decide what to do next. We call the design of this process \emph{loop engineering}, which includes context management, model selection, compaction, and termination. Existing efficiency work mainly compresses tool observations or routes selected turns to cheaper models, but the practical opportunity for these optimizations remains unclear. We analyze 600 publicly released coding-agent trajectories spanning two agent scaffolds and two language models. Median observations contain only 118--286 tokens, while the large build logs commonly used to motivate compression occur near the ninety-ninth percentile. Moreover, 98.6\% of input tokens are served from a prefix cache, so flat input pricing overestimates loop cost by roughly $7\times$ on average under near-ideal caching.

We derive and empirically validate a closed-form break-even condition for model delegation. Delegation increases cost in half of the evaluated settings, near the predicted boundary. More broadly, turn count explains 84\% of the variance in log trajectory cost, compared with 4\% for mean observation size; the most expensive trajectories have the most turns despite having the smallest observations. These associations suggest that observation-focused methods target a secondary cost factor, and identify unnecessary turns as the larger and largely unexamined opportunity, though we do not establish that turns can be removed while holding task success fixed. We release TurnCost, a tool for re-scoring trajectory corpora under alternative caching, pricing, compression, and routing policies.

\smallskip
\noindent\textit{Published in Proceedings of the 1st International Workshop on PromptOps and Vibe Coding (POVC '26), co-located with ASE 2026, Munich, Germany. DOI: \href{https://doi.org/10.1145/3843779.3844635}{10.1145/3843779.3844635}.}
\end{abstract}


\maketitle
\section{Introduction}

Agentic coding systems such as SWE-agent~\cite{sweagent} and
OpenHands~\cite{openhands}, together with their commercial descendants, now resolve
a substantial share of real GitHub issues~\cite{swebench} and are moving from
research prototypes into everyday developer tools. These systems do not answer a
question and stop. They iterate in the ReAct pattern~\cite{react}: propose an edit,
apply it, compile the project, run the tests, inspect what the tools return (the
\emph{observation}, in ReAct terminology), and decide what to do next.

Iteration introduces design decisions that lie outside any individual prompt. Which
model should handle each turn? What should remain in the transcript, and what should
be removed? When should the accumulated context be compacted, and by what mechanism?
How many turns should the agent be allowed? These are properties of the loop rather
than of a single prompt. We use the term \emph{loop engineering} for the deliberate
design and tuning of this iterative control loop as an engineering artifact in its
own right. The framing parallels prompt engineering and the broader effort to treat
developer--LLM workflows as artifacts that should be systematically designed and
measured. Prompt engineering concerns the content of one request; loop engineering
concerns the structure of a sequence of requests, actions, and observations.

\fix{The problem we address is an evaluation one. Both dominant efficiency techniques for
agentic loops are assessed against cost, but cost is computed in ways that differ by nearly
an order of magnitude depending on whether the prefix cache is priced, and on trajectories
whose lengths differ by an order of magnitude as well. Neither assumption is usually stated.
Under a cache-aware accounting that makes both explicit, we find that} current
loop-engineering work concentrates on one dimension of the cost problem while most of the
observed cost variation lies elsewhere.

\textbf{Two techniques, one target.}
Two lines of work dominate efforts to reduce the cost of agentic loops.
\emph{Context reduction} exploits the append-only structure of the transcript: because each
turn re-sends everything retained so far along with the latest tool output, that material is
processed repeatedly and cost grows faster than linearly with trajectory length. Context
reduction removes or summarizes it before the transcript is sent again~\cite{agentdiet,
opendev}. \emph{Step-level model routing} starts elsewhere: not every turn is equally
difficult. Planning a fix for a subtle failure may need a frontier model; choosing which of
forty search results to inspect may not. Routing assigns individual turns to cheaper models,
with reported savings as high as 72\%~\cite{agentrouter, swerouter, mtrouter, unoorchestra}.
Different as the mechanisms are, both target the cost of processing tool observations.
Context reduction shortens what future turns must read; routing changes the price of the
model that reads it. Neither reduces the number of turns. The size and cost of real
observations therefore bound what either can save.

\textbf{The anomaly.}
AgentDiet~\cite{agentdiet} uses a language model as a reflection module to reduce
agent trajectories. On SWE-bench Verified, it reports reductions of 39.9--59.7\% in
input tokens and 21.1--35.9\% in total cost. It also accounts for cached input
pricing, which, as we show, much of the routing literature does not.

As part of its module-selection experiment, AgentDiet evaluates a sanity-check baseline
called \emph{Delete}, which removes the observation entirely with no summarization and no
extra model call. Delete reaches $0.655$ of the unmodified baseline cost at a 58\% pass
rate; the best learned reducer reaches $0.652$ at 60\%. The learned reducer gains two
percentage points of pass rate but under half a percent of additional cost saving over a
zero-cost deletion rule despite invoking another language model and being selected from
five candidates.

Why does a learned reducer provide so little cost advantage over simply deleting the
observation? And how is that reconcilable with the far larger savings reported by routing
methods that target the cost of processing those same observations?
We explain this anomaly and show that the explanation also constrains step-level routing in
the same regime. We re-score 600 published agent trajectories spanning two scaffolds and two
language models under an exact prefix-cache simulation. The explanation turns on a simple
distributional property of real tool observations, and redirects attention to a different
part of the loop.

\subsection{Research questions}
\begin{description}[leftmargin=2.8em,style=multiline,itemsep=4pt,topsep=2pt]
\item[RQ1:] How large are tool observations in real agentic coding trajectories? Is the large build log commonly used to motivate context reduction and routing typical or rare?

\item[RQ2:] Under cache-aware pricing, by how much does flat input pricing overstate
the cost of an agentic loop?

\item[RQ3:] Does delegating observation handling to a cheaper model outperform
deterministic truncation on cost, and when does either approach pay for itself?

\item[RQ4:] Does the number of turns or the size of the observations contribute more to variation in trajectory cost?
\end{description}

\subsection{Contributions}

\begin{itemize}[leftmargin=1.1em,itemsep=2pt,topsep=2pt]
\item \textbf{The observation-size distribution} across 600 published trajectories,
two agent scaffolds, and two language models. Median observations contain 118--286
tokens, while the large-log case lies near the ninety-ninth percentile
(\S\ref{sec:obs}). This distribution explains why selective compression quickly
saturates.

\item \textbf{A cache-aware cost model and closed-form break-even condition}
(\S\ref{sec:model}). Delegation is governed by the strong model's
\emph{cached-read} price rather than its list input price, making the constraint
roughly ten times tighter than flat-price calculations suggest. Delegation increases
cost in half of our settings, near the predicted threshold
(\S\ref{sec:sweep}).

\item \textbf{A quantification of the flat-pricing error.} We find that 98.6\% of
input tokens are served from the prefix cache, making flat-price cost estimates
roughly $7.1\times$ too high on average under near-ideal caching
(\S\ref{sec:inflation}). This result applies to routing studies that use cache-unaware
accounting~\cite{agentrouter, swerouter, mtrouter}, but not to AgentDiet~\cite{agentdiet}.

\item \textbf{\fix{A decomposition of where cost variation sits.}} Turn count explains
84\% of the variance in log trajectory cost, whereas mean observation size explains
4\% (\S\ref{sec:drivers}). Every technique we evaluate operates on the dimension
associated with the smaller share of cost variation. \fix{The relationship is
associational, and \S\ref{sec:limits} separates what is definitional from what is not.}

\item \textbf{TurnCost}, an open-source re-scoring tool that recomputes the cost of published
trajectory corpora under cache-aware pricing~\cite{turncost}.
\end{itemize}

For this workshop, our argument is about the evaluation methodology used for multi-turn
developer--LLM interaction. We make both the accounting model and the evaluation regime
explicit, identify the term that determines whether an optimization pays off, and release
tooling that lets any published corpus be recomputed under an inspectable cache model.

\section{Background and Related Work}

\subsection{Context reduction}

AgentDiet~\cite{agentdiet} is the closest work to ours and our analysis builds on its
results. It reduces trajectories on SWE-bench Verified and Multi-SWE-bench using a
language-model reflection module. Its cost accounting matters here: it prices cached input
separately, states that total cost includes both input tokens with the KV-cache discount and
output tokens, and preserves cache reuse by modifying only a fixed recent portion of the
trajectory. Our criticism of cache-unaware routing costs in \S\ref{sec:model} therefore does
not apply to it. Production scaffolds use similar context-compaction strategies near context
limits~\cite{opendev}.

\subsection{Step-level model routing}

AgentRouter~\cite{agentrouter} routes each trajectory step among four model tiers with a
lightweight classifier, reporting a 71.8\% cost reduction while retaining 97.3\% of
frontier-model quality. Its trajectories contain only 4--12 steps and come from LangGraph
and CrewAI workflows rather than coding agents, and its model pool is described using
output-token prices, with input pricing unspecified.

SWE-Router~\cite{swerouter} is closer to our setting: it begins with a weaker model on
SWE-bench and escalates after $K$ exploratory turns, using flat per-inference pricing, on
the grounds that cost is model-dominated on long agentic trajectories and therefore
query independent. MTRouter~\cite{mtrouter} computes episode cost from a flat input/output
price table; it identifies the loss of prompt-cache reuse when switching models but does not
include that effect in its cost model. Uno-Orchestra~\cite{unoorchestra} routes decomposed
subtasks rather than individual turns.

TRACE-Router~\cite{tracerouter} makes one routing decision at task admission and pins all
subsequent calls to that backend, which lets the model reuse its accumulated prefix cache.
Since cached reads on a strong model may cost less than a cold prefill on a weaker one
(\S\ref{sec:model}), per-step switching can begin at a structural disadvantage. TRACE-Router
also evaluates on Terminal-Bench, a long-horizon coding benchmark closer to our regime than
short workflow trajectories.

Two patterns motivate our evaluation. First, the largest reported savings are generally
measured on 4--12-step trajectories, often outside software engineering, whereas
coding-agent trajectories are much longer and repeatedly reprocess a growing transcript.
Second, AgentRouter reports its smallest saving for code generation, 63.8\% against 76.4\%
for document analysis, consistent with the constraints we identify.

The broader literature explores related formulations, query difficulty estimation with a
variational autoencoder~\cite{daao}, reinforcement learned cost-aware routing of tool
calls~\cite{xrouter}, per-step selection between two models under an explicit
budget~\cite{budgetaware}, and contextual-bandit routing under a multi-choice knapsack
constraint~\cite{pilot}, alongside model selection in compound AI
systems~\cite{compoundai} and surveys of the wider design space~\cite{routingsurvey,
cascadesurvey}.

\textbf{Scope of our accounting claim.}
We examined AgentRouter, SWE-Router, MTRouter, AgentDiet, and TRACE-Router. AgentDiet
explicitly prices cached input. AgentRouter, SWE-Router, and MTRouter use cache-unaware
accounting: AgentRouter specifies only output-token prices, SWE-Router uses flat
per-inference pricing, and MTRouter prices from a flat input/output table despite discussing
cache loss under model switching. TRACE-Router does not report a token-level cost model, so
we make no accounting claim about it. We do not extend this claim to the wider literature
above, which we did not audit systematically. These differences make reported savings hard to
compare and motivate the cache-aware model in \S\ref{sec:model}.

\subsection{Query-level routing and token accounting}

FrugalGPT~\cite{frugalgpt}, RouteLLM~\cite{routellm}, and Hybrid LLM~\cite{hybridllm} route
independent, stateless queries rather than turns within a shared trajectory. AgentRouter
reports that applying them per step yields only 31--44\% cost reduction, because they cannot
model dependencies across turns.

Tokenomics~\cite{tokenomics} studies token use across stages of the ChatDev pipeline,
finding that code review accounts for 59.4\% of tokens and input for 53.9\% of consumption.
ChatDev is a staged pipeline rather than a loop that repeatedly re-sends an accumulating
transcript, so we analyze distributions at the turn level rather than by development stage.

\subsection{Agent--computer interfaces}

SWE-agent~\cite{sweagent} introduced the agent--computer interface (ACI), in which tools
return concise, structured observations rather than unrestricted shell output. Intended to
improve effectiveness, the design also shrinks the tool output carried through the
loop, which \S\ref{sec:whysmall} shows is central to our observation-size distribution.

\section{A Cache-Aware Cost Model}
\label{sec:model}

\fix{An agent's transcript grows over time: each turn re-sends the prior context plus the newest tool output. With prefix caching, most of that repeated context is billed at the cached-read rate, while only the new suffix is billed at full input price. Two consequences follow. Tokens added early are re-read on later turns, so cost grows faster than turn count. And a cheaper model must beat the strong model's \emph{cached} price, not its list price. This section formalizes both effects.}

\subsection{Cost structure}

Let a trajectory have $T$ turns. Write $C_t$ for the context tokens sent at turn $t$,
$o_t$ for the output tokens generated at that turn, and $r_t$ for the observation
appended after turn $t$. Let $\pin_S$, $\pout_S$ and $\pcache_S$ be the strong model's
uncached-input, output and cached-read prices, and $\pin_W$, $\pout_W$ the weak model's.
The loop is append-only, so

\begin{equation}
C_{t} = C_0 + \sum_{j<t}\bigl(o_j + r_j\bigr),
\end{equation}

which is the source of the superlinearity: a token added at turn $j$ is paid for again at
every turn after it. Under prefix caching the trajectory costs

\begin{equation}
\mathrm{Cost} = \sum_{t=1}^{T}\Bigl[\pcache_S h_t + \pin_S (C_t - h_t) + \pout_S o_t\Bigr],
\label{eq:baseline}
\end{equation}

where $h_t$ is the number of tokens at turn $t$ served from cache. Because the transcript
only ever appends, the sole genuinely new input at turn $t$ is the observation appended
after turn $t-1$, giving $h_t = C_t - r_{t-1}$.

This cache model is not an idealization introduced for convenience; it is a consequence of
how prefix caching works. A radix tree keyed on the token prefix reuses the matching
key--value entries~\cite{sglang}, so an append-only conversation hits the cache on
everything but its newest suffix. The same mechanism means switching model is expensive:
the new backend shares no prefix with the old one and must prefill the entire context from
cold.

\subsection{In-band swapping versus out-of-band delegation}

There are two very different ways to use a cheaper model inside a loop, and once caching
is taken into account their effects run in opposite directions. Under \emph{in-band
swapping}, turn $t$ is run on the weak model using the complete running transcript. The
weak model has no cache and pays $\pin_W C_t$ for a cold prefill, so the swap is only
worthwhile if $\pin_W < \pcache_S$, far more restrictive than the $\pin_W < \pin_S$ that
comparing list prices would suggest. Under \emph{out-of-band delegation}, the weak model
receives a narrow slice of the context, usually just the observation, and a short result is
inserted back into the main transcript. The strong model's cache is left intact and the
transcript is permanently shortened, so the saving accrues on every subsequent turn.

\subsection{Break-even condition}

Consider out-of-band delegation at turn $j$, replacing an observation of $r_j$ tokens with
a summary of $s_j = \sigma r_j$ tokens. The saving, realized at every subsequent turn, is
$(r_j - s_j)\,\pcache_S\,(T-j)$; the cost incurred is $\pin_W(r_j + \epsilon) + \pout_W
s_j$, where $\epsilon$ is prompt overhead. Writing $e = \epsilon/r_j$, delegation pays for
itself precisely when

\begin{equation}
T - j \;>\; \frac{\pin_W(1+e) + \pout_W \sigma}{(1-\sigma)\,\pcache_S}
\;\xrightarrow[\;e\to0\;]{\sigma\to0}\;
\frac{\pin_W}{\pcache_S}.
\label{eq:breakeven}
\end{equation}

\fix{The two sides of Eq.~\eqref{eq:breakeven} play different roles. For fixed pricing, compression, and prompt-overhead assumptions, the right-hand side is a constant; call it the break-even horizon $H$. The left-hand side, $T-j$, is the number of turns remaining when delegation occurs. A deployment can compute $H$ from its pricing assumptions and compare it with the expected remaining trajectory length at each decision.}
Delegate when the number of remaining turns is greater than the ratio of the weak model's
input price to the strong model's \emph{cached-read} price. Three effects are
counterintuitive. Observation size is absent, because it governs how much a delegation
saves rather than whether it is worth doing at all. The strong model's list price is absent,
because the relevant comparison is against its cached-read price, typically a tenth of
list. And a poor compressor, one with large $\sigma$, pushes the break-even point outwards
through both the numerator and the denominator, so the penalty for poor compression is
worse than linear.

\fix{\textbf{A worked example.} Consider a strong model priced at \$3.00 per million input tokens, cached reads at \$0.30, and a weak model at \$0.15, the 10\% and 5\% ratios used throughout. Then $H=0.15/0.30=0.5$, so one remaining turn is sufficient for delegation to break even under the limiting assumptions. Using the strong model's list price instead gives $0.15/3.00=0.05$, understating this threshold by $10\times$. Even with the correct threshold, however, delegation is not always profitable. As \S\ref{sec:sweep} shows, large observations are rare and may occur too late to amortize the extra model call, while smaller observations offer little saving per delegation.}

\section{Methodology}

\textbf{Corpus.} We use \texttt{nvidia/Open-SWE-Traces}~\cite{openswetraces}, 207,489
published agent trajectories released under CC BY 4.0. It spans two scaffolds,
OpenHands~\cite{openhands} and SWE-agent~\cite{sweagent}, crossed with two models,
Qwen3.5-122B-A10B and Minimax-M2.5, giving four cells; the instances are drawn from
SWE-rebench-V2, which spans nine programming languages. We sample 150 trajectories
from each of four parquet shards, one per cell, 600 in total. Sampling is deterministic:
we take the first 150 rows in file order with no shuffling, and the exact shard
identifiers are listed in the artifact README, so the sample reproduces byte for byte.
Because the corpus is broader than Python-only SWE-bench, our
distributional results generalize further than a single-language study, but our
absolute numbers are not directly comparable to SWE-bench-Verified-only results such as
AgentDiet's.

\textbf{Reconstruction.} Published corpora ship the full conversation but no token
accounting. Prefix-cache behavior is deterministic given an append-only
transcript, so rather than rely on whatever a provider reported, our tool,
\textbf{TurnCost}, reconstructs the
per-turn context from the message sequence, tokenizes with \texttt{o200k\_base}, and
simulates the cache exactly using $h_t = C_t - r_{t-1}$, which follows from the
append-only structure of the transcript. Tool schemas are counted, since they are
re-sent with every request and are not negligible. Reasoning tokens are billed as output
but are not carried forward into the next request, and are modeled accordingly.
\fix{For \S\ref{sec:whysmall} we additionally scan raw observation text for the two scaffolds'
truncation markers and classify the agent's next tool call as a narrowed re-query, an
identical retry, a termination, or neither; the patterns ship with the artifact.}

\textbf{Conditions.} We evaluate five configurations. \textbf{B0} is the unmodified
baseline. \textbf{A*} is in-band swapping, with the weak model receiving the full
transcript. \textbf{R1} is out-of-band delegation to a weak model.
\textbf{B2} is the same delegation performed by the strong model, which isolates the
contribution of compression from that of routing. \textbf{B3} is deterministic truncation
with no model call, the analogue of AgentDiet's \emph{Delete}. Without B2 and B3, any
saving could be attributed to compression rather than routing, and the two would be
inseparable.

\textbf{Pricing.} We normalize $\pin_S$ to \$1 per million tokens and report
ratios, so conclusions do not depend on one vendor's price list and do not expire when
prices change. Defaults are $\pcache_S = 0.1\,\pin_S$, $\pin_W = 0.05\,\pin_S$ and
$\sigma = 0.05$; all three are swept in \S\ref{sec:robust}.

\section{Results}
\label{sec:results}

\subsection{Observations are small (RQ1)}
\label{sec:obs}

\begin{figure}[t]
\centering
\includegraphics[width=\columnwidth]{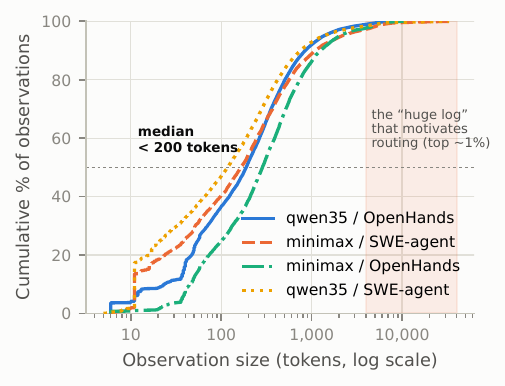}
\caption{Distribution of tool observation sizes across the four cells. The median is under
300 tokens everywhere, and the large-log case that motivates selective compression
occupies roughly the top 1\% of the distribution.}
\label{fig:cdf}
\end{figure}

\begin{table}[t]
\caption{Corpus shape and observation-size distribution. Turn counts are medians with the
interquartile range in brackets; observation sizes are percentiles over all observations
in the cell.}
\label{tab:corpus}
\small
\setlength{\tabcolsep}{3pt}
\begin{tabular}{@{}lrrrrrr@{}}
\toprule
\textbf{Model /} & \textbf{Turns} & \textbf{Ctx} & \multicolumn{4}{c}{\textbf{Observation size}} \\
\cmidrule(l){4-7}
\textbf{scaffold} & med.\ (IQR) & med. & p50 & p90 & p99 & max \\
\midrule
qwen35 / OpenHands  & 97 (85--112)   & 61{,}699 & 189 & 854 & 4{,}274 & 28{,}604 \\
minimax / SWE-agent & 71 (47--90)    & 46{,}783 & 166 & 1{,}092 & 6{,}208 & 33{,}494 \\
minimax / OpenHands & 60 (44--80)    & 45{,}395 & 286 & 1{,}258 & 5{,}194 & 12{,}854 \\
qwen35 / SWE-agent  & 134 (114--159) & 73{,}336 & 118 & 749 & 5{,}219 & 32{,}980 \\
\bottomrule
\end{tabular}
\end{table}

Figure~\ref{fig:cdf} and Table~\ref{tab:corpus} answer RQ1 and underpin everything that
follows. Median observation size ranges from 118 to 286 tokens by cell, and the ninetieth
percentile from 749 to 1,258. Very large observations exist; per-cell maxima run from
12,854 to 33,494 tokens, but they are rare enough that a scheme targeting them selectively
has almost nothing to work with.
This is why AgentDiet's Delete result is less surprising than it looks. At two or three
hundred tokens there is little semantic content for a smarter method to preserve, and
little bulk for either method to remove. The sophisticated reducer and the trivial one
converge because the object they act on is small.

The two dimensions in Table~\ref{tab:corpus} also vary independently. The cell with the
longest trajectories, qwen35/SWE-agent at a median of 134 turns, has the \emph{smallest}
median observation at 118 tokens; the cell with the largest observations,
minimax/OpenHands at 286 tokens, has the shortest trajectories at a median of 60. Scaffold
and model move trajectory length and observation size independently, a first indication of
the decomposition in \S\ref{sec:drivers}.

\subsection{The bulk comes from edits, not logs}
\label{sec:turntypes}

\begin{table}[t]
\caption{Turn types pooled across all four cells (56,164 turns), by share of turns, share
of observation tokens, and mean observation size.}
\label{tab:turntypes}
\small
\begin{tabular}{@{}lrrr@{}}
\toprule
\textbf{Turn type} & \textbf{\% of turns} & \textbf{\% of obs.\ tokens} & \textbf{Mean obs.} \\
\midrule
read / exec & 67.4\% & 43.7\% & 294 \\
edit        & 29.9\% & 51.8\% & \textbf{787} \\
terminate   & 1.6\%  & 4.5\%  & 1{,}259 \\
think       & 1.1\%  & 0.0\%  & 6 \\
\bottomrule
\end{tabular}
\end{table}

Pooling all 56,164 turns and grouping by tool gives Table~\ref{tab:turntypes}. The largest
observations do not come from reading logs; they come from \emph{edit} operations, because
an editor tool echoes the updated file context back to the agent, at a mean of 787 tokens
against 294 for read and execute turns combined.
The motivating example in this literature is therefore misplaced twice over. A scheme aimed
at very large build logs targets a tail that barely exists; a scheme aimed at what actually
dominates the transcript would have to compress editor echo, the source code the agent is
actively reasoning about, and exactly the content least safe to discard.

\subsection{Flat pricing overstates loop cost by $7.1\times$ \fix{under near-ideal caching} (RQ2)}
\label{sec:inflation}

\begin{figure}[t]
\centering
\includegraphics[width=\columnwidth]{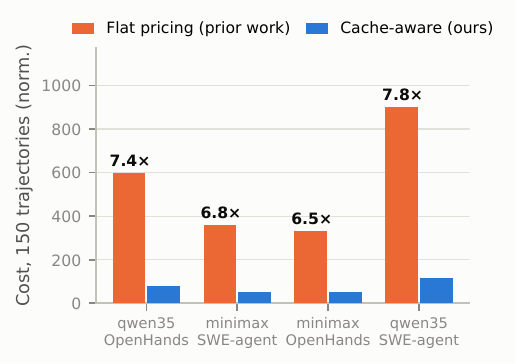}
\caption{Cost of 150 trajectories per cell under flat pricing, as used by three recent
routing papers, and under cache-aware pricing. Labels give the ratio between the two.}
\label{fig:inflation}
\end{figure}

Between 98.1\% and 99.0\% of input tokens are served from the prefix cache, so flat pricing
overstates real cost by $6.5\times$ to $7.8\times$, a mean of $7.14\times$
(Figure~\ref{fig:inflation}). \fix{This is a near-ideal-caching figure rather than a
deployment constant; \S\ref{sec:robust} gives the sensitivity ladder.} The effect is
large because these trajectories are long, at a median of 60 to 134 turns: the transcript is
re-sent every turn, and nearly all of it hits the cache every time. Savings computed against
a flat-priced baseline are therefore measured against a reference cost no real deployment
pays.
Inflation rises steadily with loop length, $5.33\times$ below 40 turns, then $6.10\times$,
$6.93\times$ and $7.44\times$ in successive strata, reaching $7.99\times$ above 130 turns.
This matters for reading the routing literature, whose largest reported savings come from
trajectories of 4 to 12 steps~\cite{agentrouter}, at the far left of that range, where flat
pricing is least inaccurate. Compression savings within the same strata are non-monotonic,
from 7.4 to 11.2\%, so we draw no conclusion from them. As \S\ref{sec:model} sets out, the
critique applies to the cache-unaware accounting in~\cite{agentrouter, swerouter, mtrouter}
and not to AgentDiet.

\subsection{Truncation and delegation are indistinguishable \fix{on cost} (RQ3)}
\label{sec:sweep}

\begin{figure*}[t]
\centering
\includegraphics[width=\textwidth]{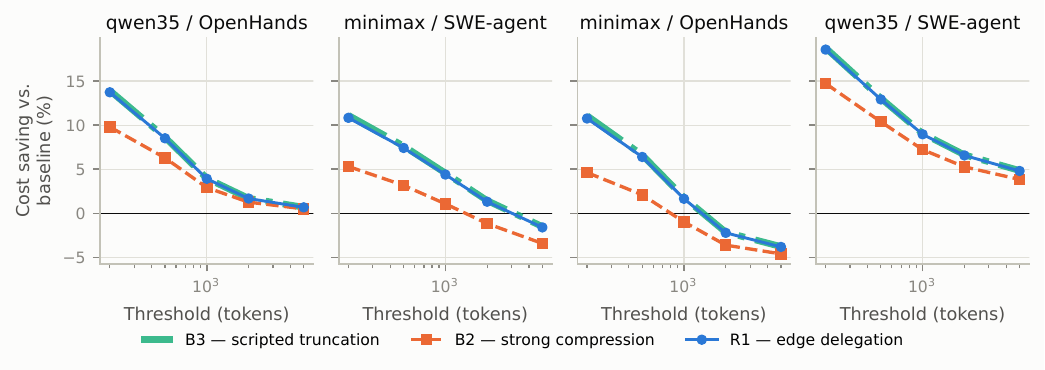}
\caption{Cost saving against the delegation threshold, for each of the four cells. B3,
deterministic truncation, is drawn thick; R1, delegation to a weak model, lies almost
exactly on top of it. B2, compression by the strong model, is consistently worse. In half
the cells the saving turns negative at conservative thresholds, as
Eq.~\eqref{eq:breakeven} predicts.}
\label{fig:sweep}
\end{figure*}

In Figure~\ref{fig:sweep} we vary the threshold above which an observation is large enough
to delegate, from 200 to 5,000 tokens. Across all twenty combinations of cell and
threshold, B3 matches or very slightly beats R1, the largest gap being 0.17 percentage
points at our default prices. This is parity, not a win for truncation: deterministic
truncation and delegation to a language model are indistinguishable on cost, and the extra
model call buys no cost advantage, just as AgentDiet found at the single point it reports.
\fix{This result concerns cost only. We do not evaluate whether learned summaries preserve information better than truncation, so we make no claim of quality equivalence. AgentDiet reports a small quality advantage, with a 60\% pass rate versus 58\% for deletion. Our claim is narrower: learned reduction should be justified by quality when it provides little additional cost saving over deterministic truncation.}

At aggressive thresholds, where nearly every observation is delegated, both methods save
11--19\%. At conservative thresholds, where only genuinely large observations are touched,
savings decline toward zero and turn negative in two of the four cells. Compressing a few
large observations late in a trajectory does not repay its own cost, because too few turns
remain to spread the saving over. This is Eq.~\eqref{eq:breakeven} visible in the data, and
it generalizes AgentDiet's single reported point across the whole threshold space.
Price ratios are not the limiting factor. At realistic ratios, weak-model input at 2--10\%
of the strong model's list price, cached reads at about 10\% of the same list price---Eq.~\eqref{eq:breakeven}
puts break-even at under two remaining turns, so delegation should repay itself almost
immediately. It does not, and Figure~\ref{fig:cdf} explains why: the binding constraint is
observation size, not model price.

\subsection{In-band swapping sits on a knife-edge}
\label{sec:astar}

In condition A*, the model is swapped for a whole turn, so the weak model receives the
complete transcript and starts cold. As \S\ref{sec:model} predicts, this helps if and only
if $\pin_W < \pcache_S$. At our defaults, $\pin_W = 0.05$ and $\pcache_S = 0.10$, the
inequality holds and A* does save: 14.7\% at a 200-token threshold, falling to 0.4\% at
5,000, and at the most aggressive threshold it beats R1, 14.7\% against 13.5\%.
The point is the phase boundary rather than the technique. A weak model priced at 20\% of
the strong model's list price appears to offer an 80\% saving under flat accounting but is a
net loss once caching is counted. Profit and loss are separated by a factor of two, and list
prices alone give no warning that one is standing on that boundary.

\subsection{Turn count, not observation size, drives cost (RQ4)}
\label{sec:drivers}

\begin{figure}[t]
\centering
\includegraphics[width=\columnwidth]{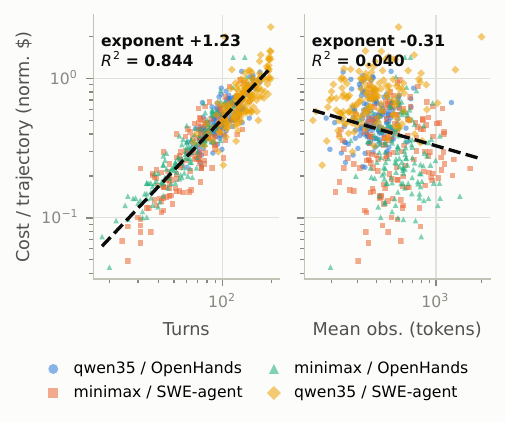}
\caption{Per-trajectory cost against the two candidate drivers, on shared axes. Turn
count, left, produces a tight power law with an exponent above one. Mean observation size,
right, produces a shapeless cloud with a slightly negative slope. Both panels contain the
same 600 trajectories.}
\label{fig:drivers}
\end{figure}

We regressed per-trajectory cost on turn
count and on mean observation size, both on a log--log scale, across all 600 trajectories
(Figure~\ref{fig:drivers}). Turn count alone gives $R^2 = 0.844$ with an exponent of 1.23;
mean observation size alone gives $R^2 = 0.040$ with an exponent of $-0.31$; together they
give $R^2 = 0.945$. Adding turn count to a model that already contains mean observation
size raises $R^2$ by 0.905; adding mean observation size to a model that already contains
turn count raises it by 0.101. The turn-count exponent exceeds one, as the append-only
re-send structure predicts, because each turn re-reads all those before it.
Sorting trajectories into deciles by turn count makes the point plainly. From the first
decile to the tenth, median turn count rises from 36 to 168 and median cost rises sevenfold,
from $0.138$ to $0.978$, while median observation size \emph{falls} from 655 to 334 tokens.
The most expensive trajectories in this corpus have the most turns and, at the same time,
the smallest observations. Cost per turn is comparatively stable, with a median of $0.005$
and a coefficient of variation of 0.28, whereas cost per observation token is less stable,
at 0.36.

\fix{These are associations, not causal effects. Because cost is summed over turns, some relationship between $T$ and cost is expected by construction; \S\ref{sec:limits} separates this mechanical effect from the empirical findings. We therefore do not claim that reducing turns would yield the savings implied by the regression or preserve task success.} What the decomposition does establish is where the variation sits, and
that is the paper's positive result for loop engineering: every technique we
examine (routing, learned reduction, truncation) operates on the observation axis, which
carries 4\% of the explained variance in cost, while the turn axis, which carries 84\%, is
left untouched by all of them.

\subsection{Ceiling and robustness}
\label{sec:robust}
Observations contribute 66--74\% of transcript growth across the four cells; the
remainder is agent output and cannot be reduced by observation compression. An oracle
that compresses every observation to 5\% of its original size at no cost and without
quality loss saves 47.8--60.3\%, depending on the cell. This is the oracle ceiling for
our assumed compression ratio, $\sigma=0.05$, rather than an absolute ceiling for all
possible compression. Approaching it requires reducing observations throughout the
trajectory, which makes a separate model call for every observation increasingly
expensive and helps explain why deterministic reduction remains competitive.

\textbf{Imperfect caching.} The $7.1\times$ figure assumes near-ideal prefix caching and is
sensitive to that assumption: it falls to $4.0\times$ at 90\% cache efficiency,
$2.4\times$ at 75\%, and $1.45\times$ at 50\%. We therefore treat $7.1\times$ as a value
attained under near-ideal caching rather than a general constant, and ask that it be cited
with that caveat attached. The design comparison is unaffected: the B3--R1 gap stays
non-negative at every efficiency level tested, so the equivalence of truncation and
delegation does not depend on assuming a perfect cache.

\textbf{Vendor pricing.} Varying $\pin_W/\pin_S$ over $[0.02, 0.20]$ and
$\pcache_S/\pin_S$ over $[0.05, 0.50]$, the B3--R1 gap stays non-negative in all sixteen
cases, widening to $+1.0$ points exactly where theory predicts: where the weak model is
expensive relative to cached reads.

\subsection{Why the opportunity is small}
\label{sec:whysmall}

The obvious explanation for small observations is that scaffolds truncate long ones before
the agent sees them. That is not the reason. SWE-agent's \texttt{max\_observation\_length}
defaults to 100,000 characters, roughly 25,000 tokens, consistent with the maxima we
observe. Truncation does bind at the extreme tail, but it sits two orders of magnitude
above the median: it caps the tail, it does not set the median.
What sets the median is the design of the agent--computer interface. SWE-agent's central
contribution was a set of purpose-built tools returning concise, structured output rather
than raw shell dumps~\cite{sweagent}. Because viewing a region of a file, running a specific
test, or listing search results each yields only a few hundred tokens by construction, the
compression opportunity that context reduction and step-level routing compete for had
largely been removed by interface design before either method was applied. That is good news
for the field and bad news for these techniques, and the single most useful thing a loop
engineer can know before choosing where to spend effort.

\fix{\textbf{What happens when the interface truncates.}}
\fix{Truncation provides a useful view of how agents respond when tool output is cut. Across 56,164 observations, truncation markers appear 518 times (0.92\%), spanning 367 of 600 trajectories. In 468 cases, the scaffold gives recovery guidance, such as narrowing the query with \texttt{grep -n}; the remaining 50 provide only a truncation marker. After guided truncation, 76.9\% of next actions narrow the query and exactly one, 0.2\%, terminates the trajectory, compared with 44.0\% narrowing after bare truncation; no agent repeats the identical command. These traces do not show whether the missing content was recovered correctly, but they suggest that explicit recovery guidance helps. A deletion policy should therefore indicate that content was removed and provide a way to retrieve it.}

\section{Implications for Loop Engineering}

Three recommendations follow. \emph{Price the cache before optimizing anything.} If your
workload resembles those we study, 98--99\% of your input tokens are cache hits and your
real cost is several times below a list-price estimate; an optimization tuned against the
list-price figure will underdeliver in production by the same factor. \emph{Use
Eq.~\eqref{eq:breakeven} as a deployment rule.} Delegate an observation only when the
remaining turns exceed the limiting break-even horizon $\pin_W / \pcache_S$, which at the tariff in
\S\ref{sec:model} is under one turn. If delegation still fails to break even there, the
price gap is not the problem, and no amount of router tuning will fix it.
\emph{Always compare against a truncation baseline.} A method that does not beat deleting
the observation is buying quality, not cost, and should be justified on quality
grounds: a cheap comparison, and a decisive one on our data.

The underlying question is where to spend effort. Selective per-observation intervention
acts on the axis carrying 4\% of the variance in cost; turn reduction acts on the axis
carrying 84\%, and removing a turn eliminates not only that turn's cost but every
subsequent re-read of it. We are not aware of work targeting redundant turns directly in
coding agents. \fix{Our evidence for this is associational rather than experimental
(\S\ref{sec:limits}), so we frame it as a direction rather than a guaranteed saving: the
open question is why an agent needs a hundred turns to fix one bug, and whether those turns
can be removed without costing task success. That question is unanswered, and it is a more
promising place to look than shaving tokens off observations that were already small.}

\section{Limitations}
\label{sec:limits}

\textbf{Corpus provenance.} The corpus was assembled as distillation training data, and its
quality filters are not neutral for us. It \emph{excludes} trajectories ``where the agent
reached the maximum iteration limit or the trajectory was prematurely terminated by the
harness, regardless of whether the resulting patch resolved the issue''~\cite{openswetraces},
along with runs producing empty patches or altering the test suite. This truncates the right
tail of the turn-count distribution, with opposite effects on our two results. For flat
pricing the bias is \textbf{in our favor}: inflation grows with trajectory length, so
dropping the longest runs makes our $7.1\times$ conservative. For observation size it is
\textbf{adverse and unbounded}: an agent that re-reads a large log until it exhausts its
iteration budget is precisely what was excluded, so the true distribution may have a heavier
tail than Figure~\ref{fig:cdf} shows. We therefore treat the medians as indicative rather
than exact.

\textbf{The turn-count association is partly mechanical.} Cost is computed as a sum over
turns, so some relationship between $T$ and cost holds by construction, and $R^2 = 0.844$ is
not a causal finding. Two features are not definitional. The exponent exceeds one, at 1.23,
so cost grows faster than turn count (the re-read effect, not merely more calls). And mean
observation size is not just a weaker predictor but a slightly negative one, at $-0.31$:
longer, costlier trajectories carry smaller observations. The two axes are therefore not two
views of one quantity, and the practical claim survives the caveat.

\textbf{Model and tokenizer scope.}
We study open-weight models rather than frontier proprietary agents. Although the
patterns are consistent across both models and scaffolds, frontier systems may behave
differently. We also use \texttt{o200k\_base} instead of each model's native tokenizer,
so absolute token counts may differ by 10--20\%. Using one tokenizer consistently makes
the comparisons less sensitive to this mismatch, but does not eliminate it entirely,
especially if the models tokenize code and tool output differently.

\textbf{Compression quality is not measured.}
We set $\sigma=0.05$ but do not evaluate whether semantic summaries preserve useful
information better than deterministic truncation, so the parity result in \S\ref{sec:sweep}
concerns cost and not quality; AgentDiet's 60\% against 58\% pass rate is the only quality
evidence we have. \fix{The truncation measurements in \S\ref{sec:whysmall} bear on what an agent does when
material is withheld, but they are observational and do not establish that a deletion policy
is harmless.} Evaluating the tradeoff across the full threshold range is a natural next step
that trajectory re-scoring cannot reach.

\textbf{Cache model.}
Beyond the efficiency sweep in \S\ref{sec:robust}, real providers round cache entries to
block boundaries, evict them after a time-to-live, and lose reuse when requests move between
backends, none of which we model. The relative comparison among methods is robust to this;
the magnitude of the accounting gap is not.

\section{Conclusion}

The saturation of context reduction in coding agents has a simple, measurable cause: tool
observations are small. Across two scaffolds and two models the median observation is
118--286 tokens, and the large build log that motivated the technique sits in the top
percentile. That is why a trivial delete baseline matches a learned reducer, and it also
bounds what step-level routing can achieve in this regime. Separately, flat input pricing
overstates the cost of an agentic loop by roughly $7.1\times$ on average \fix{under
near-ideal caching}, so routing savings reported against such a baseline cannot be read directly and
are not comparable across papers that account differently.

The most interesting finding is the positive one. Turn count accounts for 84\% of the
variation in log trajectory cost and observation size for 4\%, yet both context reduction
and step-level routing act on the second axis. The first is largely unexamined, and
removing a redundant turn also removes every later re-read of it. \fix{Whether those turns
can be removed without costing task success is an open question, and answering it is where
we expect loop engineering to gain most.}

\section*{Data-Availability Statement}

TurnCost, the analysis code, and the derived results that regenerate every table and
figure in this paper are archived at Zenodo~\cite{turncost}. The trajectory corpus is publicly available~\cite{openswetraces} and is not redistributed.

\bibliographystyle{ACM-Reference-Format}
\bibliography{refs}
\end{document}